\documentstyle[epsf]{article}

 {\par\noindent{\underline{Proof} \quad}}{\hfill$\Box$\bigskip}
 {\par\noindent{\underline{Proof} of the theorem\quad}}{\hfill$\Box$\bigskip}
 {\par\smallskip\noindent{\underline{{\it Remark}} \quad}}{\par\smallskip}
 {\par\smallskip\noindent{\underline{{\it Fact}} \quad}}{\par\smallskip}
 {\par\smallskip\noindent{\underline{{\it Example}} \quad}}{\par\smallskip}
 {\par\smallskip\noindent{{\it Assumotion} \quad}}{\par\smallskip}
 {\par\smallskip\noindent{{\it Condition} \quad}}{\par\smallskip}

\begin{document}
% commands
%\newcommand{\rgl}{\rangle}
\title{
Non-adiabatically detecting the geometric phase of the macroscopic 
quantum state with symmetric SQUID }
\author{Wang Xiangbin, Matsumoto Keiji \\
        Imai Quantum Computation and Information project, ERATO, Japan Sci. and Tech. Corp.\\
Daini Hongo White Bldg. 201, 5-28-3, Hongo, Bunkyo, Tokyo 113-0033, Japan}

\maketitle \begin{abstract}
\begin{bf} 
We propose a non-adiabatic scheme to detect the the geometric phase for the 
macroscopic state
with the Josephson junction system. After extending the scheme to the two qubit 
cases, we provide
a type of non-adiabatic geometric C-NOT gate which plays an important role 
in quantum computation.
\end{bf}\end{abstract}

Testing quantum mechanics laws in macroscopic level is an important 
and interesting topic\cite{naka,jona}. Geometric 
phase\cite{panch, berry, aha} plays
an important role in quantum interferometry and many other disciplines. 
It should be interesting to detect the geometric phase in macroscopic 
quantum state.  Recently, after the proposal of geometric C-NOT gate\cite{ekert}  with NMR,
there
is a proposal\cite{vedral}    to adiabatically
detect the Berry phase 
with Josephson junction of asymmetric SQUID, and also, 
 to make a fault tolerate C-NOT gate through
condition Berry phase shift\cite{ vedral}. However, one should overcome two drawbacks 
in the suggestions in doing the geometric quantum computation or detecting the Berry phase only. 
One is the adiabatic condition which
makes such gate not practical, the other is the extra operation to eliminate the dynamic phase. 
This extra operation seriously weakens the fault tolerate property.
In this letter, we give a 
simple scheme to detect the geometric phase or realize the C-NOT gate non-adiabatically on the $0$ dynamic phase evolution curve
with symmetric SQUID. 
By this scheme, the detection or the realization can be done faster and more easily.
 In our scheme, the above mentioned two drawbacks of previous suggestions are removed. 
We believe our scheme has led the idea of geometric C-NOT gate much more practical than before.

For the two level system, Geometric phase is equal to half of the solid angle subtended
by the area in the Bloch sphere enclosed by the closed evolution loop of the eigenstate.
Recently\cite{ekert}, it was proposed to make a fault tolerate C-NOT gate using Berry$'s$ phase, i.e.,
adiabatic and cyclic geometric phase. It is well known\cite{bar} that, with single qubit rotation operation, C-NOT
gate can be used to carry out all computation tasks.
The experimental results with systematic errors were obtained on NMR\cite{ekert}. Very recently, there is a new proposal
to test the geometric phase and the conditional geometric phase shift with asymmetric SQUID. 
This is rather significant because 
testing the geometric phase on macroscopic quantum state itself is important 
in quantum mechanics, and also, a C-NOT gate with Josephson junction is undoubtedly a real
$quantum$ C-NOT gate. However, both proposals\cite{ekert, vedral} rely on the adiabatic operations. This is 
a bit unrealistic because the macroscopic state of Josephson junction dephases fast\cite{naka}.
The dephasing effects may distort the laboratory observation of the Berry phase seriously.
On the other hand, for the purpose of quantum computation, any quantum gate should be run as fast
as possible. Besides the adiabatic condition, both of the previous proposals take extra operation to
eliminate the dynamic phase. This extra operation is unwanted for a fault tolerate C-NOT gate because if we can not 
eliminate
the dynamic phase exactly, the fault tolerate property is weakened. 
For these reasons one is tempted to set up a new scheme which does not rely on the adiabatic condition and
which does not involve any dynamic phase in the whole process.

Geometric phase also exists in non-adiabatic process.     
It was  shown by Aharonov and Anandan\cite{aha} that the
geometric phase is only dependent on the area enclosed by the loop of the 
state on the Bloch sphere. In non-adiabatic case, the path of the state evolution in general is 
different from the path of the parameters in the Hamiltonian. 
The external field 
need not always follow the evolution path of the state like that in adiabatic case. 
So it is possible to let the external field
perpendicular to the evolution path instantaneously so that there is no dynamic phase 
involved in the whole process.
             
\begin{figure}
\begin{center}
\epsffile{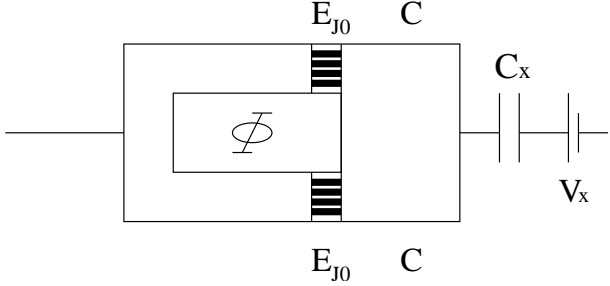}
\end{center}
\caption{ {\bf SQUID with symmetric Josephson junctions.} It consists of a superconducting box
formed by a symmetric SQUID, pierced by the magnetic voltage $\Phi$. $V_x$ is the applied voltage,
which determines the offset charge $n_x$. The device operates in the charge regime, i.e. $E_{J0}<<E_{ch}$. 
}
\label{jo}    \end{figure}

\begin{figure}
\begin{center}
\epsffile{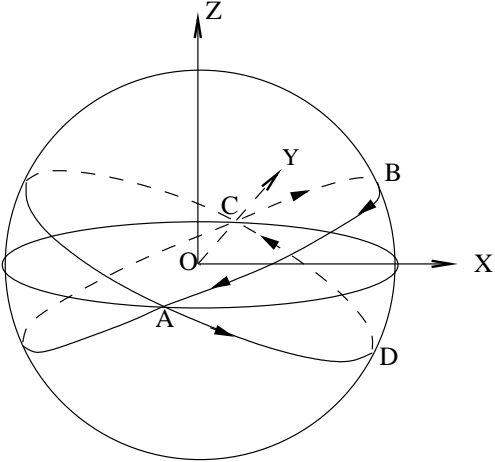}
\end{center}
\caption{ {\bf A scheme for geometric phase detection on symmetric SQUID.} 
By suddenly changing the paramecters of $n_x$ and $\Phi$, we can get the sudden fictitious field in $x-z$
plane in directions perpendicular to geodesic plane CBA and ADC, respectively. 
The fictitious field will create the evolution path of CBADC.
The angle between geodesic plane ABC and the equator is $\theta$. The solid angle subtended by  the area of
ABCDA   is $4\pi\theta$ }
\label{single}    \end{figure}

Consider a superconducting electron box formed by a $symmetric$ SQUID( see fig. \ref{jo}), pieced 
by a magnetic flux
$\Phi$ and with an applied gate voltage $V_x$. The device is operated in the charging regime, i.e.,
the Josephson couplings $E_{J0}$
 are much smaller than the charging energy $E_{ch}$. Also a 
temperature much lower than the Josephson coupling is assumed. 
The Hamiltonian for this system
is\cite{vedral, mak, shnirman}
$$
H=E_{ch}(n-n_x)^2-E_J(\Phi)\cos\chi,
$$
$E_J(\Phi)=2E_{J0}\cos\left(\pi\frac{\Phi}{\Phi_0}\right)$, $E_{ch}$ is the charging energy and $n_x$
can be tuned by the applied voltage $V_x$ through $V_x=2en_x/C_x$( see figure \ref{jo}). The phase difference
across the junction $\chi$ and the cooper pain number $n$ canonical conjugate variables $[\theta,n]=i.$ 
$\Phi_0=h/2e$ is the quantum of flux.
So here $n_x$ and $\Phi$ can be tuned externally. As it was
pointed out earlier\cite{vedral, mak, shnirman}, when $n_x$ is around $1/2$, only two charge eigenstates
$|0>$ and $|1>$ are important. The effective Hamiltonian in the computational two dimensional
Hilbert space is $H=-\frac{1}{2}\bf{B}\cdot \sigma$, where the fictitious field 
${\bf B}=[E_J, 0, E_{ch}(1-2n_x)]$ and $\sigma$ are Pauli matrices. In this picture, the state
of $0$ cooper pair($|0>$) or $1$ cooper pair($|1>$) is expressed in $|\downarrow>$ or $|\uparrow>$ respectively,
where  states $|\downarrow>$ and $|\uparrow>$   are eigenstates of Pauli matrix $\sigma_z$.
Suppose initially $\Phi=\pi/2$ and $n_x=0$, the initial state is $|\psi_0>=|0>=|\downarrow>$.
To avoid the dynamic phase, we select the geodesic curves in the time evolution. We can use the following
scheme(see fig. \ref{single}). \\
1. We suddenly change $n_x$(Note that $n_x$ can be tuned through $V_x$.) from $0$ to 
$n_0=\frac{1}{2}(1-\delta), \delta>0$.  \\
2. Wait for a time 
\begin{eqnarray}
\tau=\frac{\pi}{\sqrt{(E_{ch}\delta)^2+(2E_{J0})^2}}.
\end{eqnarray}
After this the state is rotated along geodesic curve CBA for an angle $\pi$. See fig. \ref{single}.\\
3. We suddenly change $n_x$ to $\frac{1}{2}+\delta$, and wait for a time $\tau$ again.
 The state is rotated along geodesic curve ADC for an angle $\pi$ again. See fig. \ref{single}.\\
4. Do the meassurement to detect the interference effects. 

In the above operation, 
the time evolution curve will enclose an area in Bloch sphere. The eigenstate basis $|\pm>$
, which are eigenstates of Pauli matrix $\sigma_y$ will acquire a geometric phase 
of $\pm \gamma=\pm 2\pi\theta$,  
respectively. If the initial state is $|\psi(0)>=|0>$, at time $2\tau$ it is
$$
|\psi(2\tau)>=U(2\tau) |0>,
$$
where $U(2\tau)$ is the time evolution operator from time $0$ to $2\tau$. It has the property
$$
U(2\tau)|\pm>=e^{\pm i\gamma}|\pm>.
$$
Using the relation
$$
|0>= -\frac{i}{\sqrt 2} (|+>-|->)
$$
we have
\begin{eqnarray}
|\psi(2\tau)>=-\frac{i}{\sqrt 2} U(2\tau)(|+>+|->)=-\sin\gamma |\uparrow> -\cos \gamma |\downarrow>
\end{eqnarray}
In general, suppose the initial state is $\Psi(0)>=\alpha |\uparrow>+\beta|\downarrow>$ with 
constraint of $|\alpha|^2+|\beta|^2=1$, we have
\begin{eqnarray}
|\Psi(2\tau)>=(\alpha\cos\gamma-\beta\sin\gamma) |\uparrow>-(\alpha\sin\gamma+\beta\cos\gamma)|\downarrow>.
\end{eqnarray}
We can see that, the non-zero $\gamma$ phase here will cause detectable interference effects.
Specifically, if we choose appropriate $\delta$ value so that $\gamma=\frac{\pi}{2}$, we can observe the 
flip to initial state. This property can be  potentially useful to realize the fault tolerate NOT gate in
geometric quantum computation\cite{ekert, vedral}.  
 Equivalently,
we can take rotations to the state on 
Bloch sphere around $x-$axis for the same angle $\theta$ instead of changing $n_x$ in the scheme. We
do so in the following for the conditional geometric phase shift. 

So far we have a simple scheme to non-adiabatically 
detect the geometric phase for the macroscopic quantum state of
symmetric SQUID. 
Scince it is operated non-adiabatically, the total time needed here should be comparable to one cycle
in the experiment of ref\cite{naka}. That is to say, much shorter than the decoherence time.
We believe the experiment of our scheme can be done with the same set up in ref\cite{naka}.
We have chosen the zero dynamic phase 
path\cite{suter}. We need not have to make additional evolution loop 
to cancel the dynamic phase and preserve the geometric phase, which is obviously too

\begin{figure}
\begin{center}
\epsffile{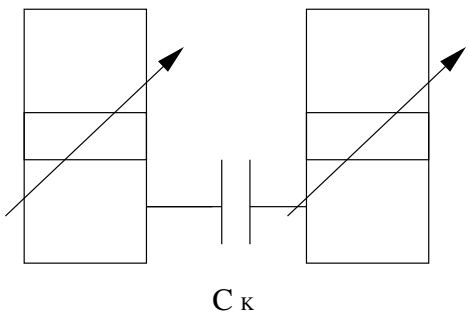}
\end{center}
\caption{ {\bf The capasitively coupled SQUIDs.} 
}
\label{couple1}    \end{figure}

The above scheme for non-adiabatic detection of geometric phase to single qubit can be easily developed 
to the two  qubit system to make a 
C-NOT gate with symmetric SQUID. Suppose we have 
two capasitively coupled symmetric SQUIDs( see fig\ref{couple1}) $1$ and $2$ 
with same $E_{J0}$ and same $E_{ch}$. 
For simplicity we call them as qubit 1 and qubit 2, respectively. We will use subscripts $1$ and $2$
to indicate the corresponding qubits.
For qubit $1$( control bit) we set $\Phi_1=\pi/2$ and $n_{x,1}=0$ during the whole process. For qubit
$2$( target bit), we set $\Phi_2=\Phi_0/2$ and $n_{x_2}=1/2$ initially.  
In quantum computation, the state for qubit $1$ and $2$ can be either $|0>=|\uparrow>$ or 
$|1>=|\downarrow>$. 
The $weak$ interacting Hamiltonian is 
$$
H_I=\Delta(n_1-n_{x,1})(n_2-n_{x,2})
$$
the constant $\Delta$ is the charging energy that derives from the capacitive coupling. We assume
$\Delta$ is much smaller than  $E_{ch}$. In our setting, the Hamiltonian reads
\begin{eqnarray}
H_I=\frac{1}{2}\Delta \sigma_{1z}\cdot \sigma_{2z}.
\end{eqnarray}         
Through our setting, to qubit  $2$ the interacting Hamiltonian is an effective
conditional Hamiltonian dependent on the specific state of qubit $1$. 
Explicitly, it is $\frac{1}{2}\Delta \sigma_z$ if state of qubit $1$ is 
$|1>=|\uparrow>$, and {\it it is $0$} if state of qubit $1$ is $|0>=|\downarrow>$.

\begin{figure}
\begin{center}
\epsffile{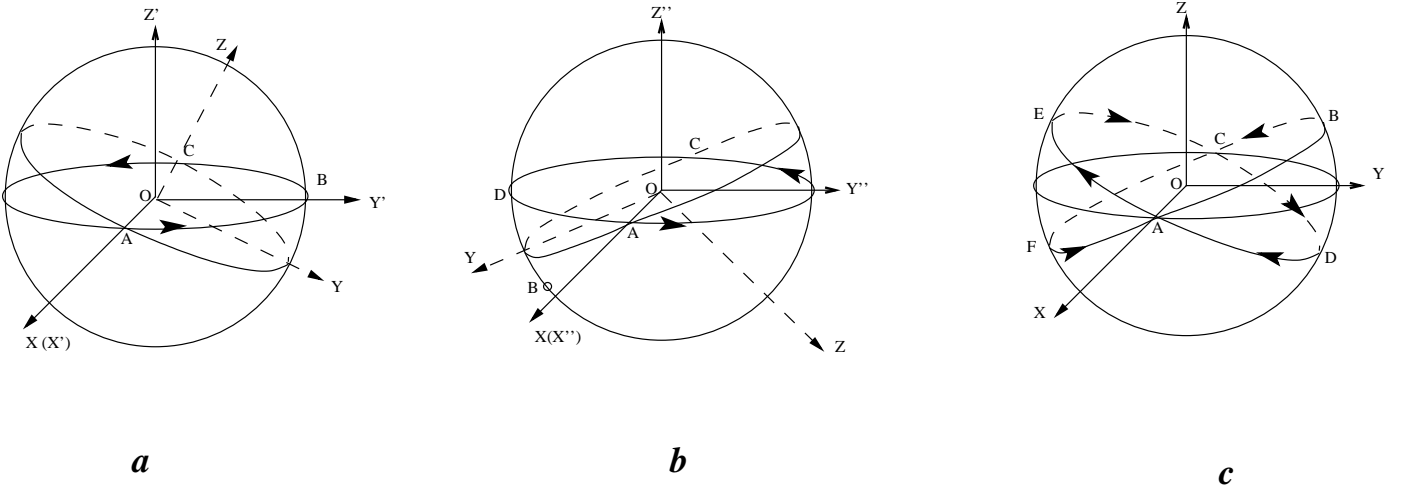}
\end{center}
\caption{ {\bf Non-adiabatic conditional geometric phase shift acquired through $0$ dynamic phase
evolution path}. These are pictures for the time evolution on Bloch sphere of qubit $2$ $only$ in the
case that
qubit $1$ is $|\uparrow>$. If qubit $1$ is $|\downarrow>$, there is no net change to qubit $2$.
   Picture $a$ shows that after the Bloch sphere is rotated around
$x-$axis for $-\theta$ angle, the interaction Hamiltonian will rotate 
the 
Bloch sphere around $z'-$axis. At the time it completes $ \pi$  rotation,
i.e. $\tau=\pi/\Delta$
we rotate the bloch sphere around $x-$axis again for an angle of $-(\pi-2\theta)$,
 then we 
get  picture ${\bf b}$. In picture ${\bf b}$ the sphere is rotated around
$z''-$axis by the interaction Hamiltonian. Note that point $B$ in picture
$b$ has changed its position now. The geodesic cure CBA is not drawn
in picture $b$. After  time $\tau$ we
rotate the qubit $2$ around $x-$axis for an angle
of $\pi-\theta$.} Picture $c$  shows the whole evolution path on the Bloch sphere. point $A$ evolves
along closed curve ABCDA, a geometric phase $\gamma=-2\theta$ is acquired. Point $C$ evolves along the 
loop   CFAEC, a geometric phase $-\gamma=2\theta$ is acquired.
\label{cnot}    \end{figure}

We first 
rotate qubit 2 around  $x-$axis for angle $-\theta$ see fig.( \ref{cnot}). 
Note that $E_{ch}$ is much larger than  $\Delta$,   so
the state of qubit $1$ is (almost) not affected by any operation in qubit $2$ in the whole process. 
The interaction Hamiltonian will create an evolution path on the geodesic circle ABC( see fig. \ref{cnot}).
After time 
$\tau=\pi/\Delta$,  we rotate qubit $2$ around $x-$axis for another angle $-(\pi-2\theta)$. 
Again wait for a  time $\tau$.
Then rotate qubit $2$ around $x-$axis for angle $\pi-\theta$ to let the Bloch sphere back to the original
one. After the above operation, if qubit $1$ is on state $|\uparrow>$,
an evolution path of ABCDA on the Bloch sphere is produced; if qubit $1$ is on state 
$|\downarrow>$, nothing happens to qubit $2$.

In the scheme we have used the rotation operation to qubit $2$, around $x-$axis. This can be done
by suddenly changing $\Phi_1$ to $0$ and $n_{x,1}=\frac{\Delta}{E_{ch}}+\frac{1}{2}$ 
to qubit $2$. The sudden change of $n_{x,1}$  is  to cancel the effects from the interaction
Hamiltonian if qubit $1$ is on state $|\uparrow>$, thus an exact $-\theta$ rotation around $x-$axis
is guaranteed. Everytime after the rotation task around $x-$axis is completed, we always set 
the parameters of 
 $\Phi_1$ and $n_{x,1}$ back to $\pi/2$ and $1/2$, respectively.
Note we need not worry about the case of $|\downarrow>$
state for qubit $1$. Because the interaction is $0$ in the case, qubit $2$ is just rotated around
an axis which is slightly different from $x-$axis and is then rotated back around the same axis.
There is still no net effect to qubit $2$ in the operation. 

Thus for qubit $2$, the final state is changed by the unitary transformation $U(2\tau)$ in the following way
\begin{eqnarray}
U(2\tau)\left(\begin{array}{c}|\downarrow>|\downarrow>\\
|\downarrow>|\uparrow>\\ |\uparrow>|\downarrow>\\ |\uparrow>|\uparrow>\end{array}\right)=
\left(\begin{array}{cc}I&0\\0&M\end{array}\right) \left(\begin{array}{c}|\downarrow>|\downarrow>\\
|\downarrow>|\uparrow>\\ |\uparrow>|\downarrow>\\ |\uparrow>|\uparrow>\end{array}\right)
\end{eqnarray}
where $M=\left(\begin{array}{cc}\cos \gamma(\theta)& i\sin\gamma(\theta)\\
 \\ i\sin\gamma(\theta) & \cos\gamma(\theta)
\end{array}\right)$, $\gamma(\theta)$ the geometric phase acquired for initial state $|X,+>$(point $A$ in
fig.\ref{cnot}) and $\gamma(\theta)=-2\theta$. 
We see $|\gamma(\theta)|=\pi/4$ makes a C-NOT gate here( see fig\ref{cnot}). This corresponds to 
$\theta=\pi/8$.

Thus, our sceme can be used to make a geometric C-NOT gate, which is fault tolerate 
to certain types of errors\cite{ekert, duan}. Since the adiabatic condition is removed, the total operation
time needed here should be comparable to that of a normal C-NOT gate. Therefore we believe our scheme has led the 
idea of geometric quantum computation much closer to the practical application. The various parameters of a normal
C-NOT gate are listed in ref\cite{bar,mak,mak1}. 
It has been estimated there that the operation time can be much shorter
than the decoherence time. Instead of the inductively coupled system, here we have used the capasitively
coupled system, however, this is not an essential mordification.  
In the real experiment, one has to have certain readout facility. 
The single eletron transistor can be a good candidate\cite{mak1}. 

{\bf Acknowledgement:} We thank Prof Imai for support. We thank Dr Y. Nakamura(NEC) for fruitful discussions.

\newpage
\newpage
\newpage
\newpage
\end{document}